\documentclass[useAMS,usenatbib]{mn2e}

\usepackage{graphicx}
\usepackage{multirow}
\usepackage{amssymb}
\usepackage{aas_macros}

\def \sdss     {{\it SDSS}}
\def \wise     {{\it WISE}}

\def \mum {\ensuremath{\,\mathrm{\mu m}}}

\def \oiii {\ensuremath{\mathrm{[O}${\sc iii}$\mathrm{]}}}

\def \sii {\ensuremath{\mathrm{[S}${\sc ii}$\mathrm{]}}}
\def \ha {\ensuremath{\mathrm{H\alpha}}}
\def \hb {\ensuremath{\mathrm{H\beta}}}

\def \mbh {\ensuremath{M_\mathrm{BH}}}

\def \loiii {\ensuremath{L$\oiii$ }}
\def \lwtwo {\ensuremath{L_{4.6\,\mathrm{\mu m}}}}
\def \lwfour {\ensuremath{L_{22\,\mathrm{\mu m}}}}
\def \dfour {\ensuremath{D_\mathrm{n}(4000)}}

\def \eddpar {\ensuremath{\loiii/\mbh}}
\def \wtwoeddpar {\ensuremath{\lwtwo/\mbh}}
\def \wfoureddpar {\ensuremath{\lwfour/\mbh}}
\def \wtworat {\ensuremath{\lwtwo/\loiii}}
\def \wfourrat {\ensuremath{\lwfour/\loiii}}

\title[The Nature of Obscuration in AGN]{The Nature of Obscuration in AGN: II. Insights from Clustering Properties}

\author[L.~Shao et al.]{Li Shao,$^1$\thanks{E-mail: lishao@mpa-garching.mpg.de} Cheng Li$^2$, Guinevere Kauffmann$^1$ and Jing Wang$^{3}$\\
$^1$ Max Planck Institute for Astrophysics, Karl-Schwarzschild-Str. 1, Garching, 85748, Germany\\
$^2$ Partner Group of the Max Planck Institute for Astrophysics and Key Lab for Research in Galaxies and Cosmology,\\
  Shanghai Astronomical Observatory, Nandan Road 80, Shanghai, 200030, China\\
$^3$ CSIRO Astronomy \& Space Science, Australia Telescope National Facility, PO Box 76, Epping, NSW 1710, Australia\\}

\begin{document}

\date{Accepted on 9 December 2014}

\pagerange{\pageref{firstpage}--\pageref{lastpage}} \pubyear{2014}

\maketitle

\label{firstpage}

\begin{abstract}
  Based on large optical and mid-infrared (IR) surveys, we investigate
  the relation between nuclear activity in local Seyfert 2 galaxies
  and galaxy interactions using a statistical neighbour counting
  technique. At the same level of host galaxy star formation (SF), we
  find that active galactic nuclei (AGNs) with stronger \oiii\ emission
  lines do not show an excess of near neighbours, while AGNs with
  stronger mid-IR emission do have more near neighbours within a
  projected distance of 100 kpc. The excess neighbour count
  increases with decreasing projected radius. These results 
  suggest a phase of torus formation during galaxy interactions.
\end{abstract}


\section{Introduction}
\label{sec:intro}

An empirical tight correlation between the stellar mass of a galaxy
bulge and the mass of its central supermassive black hole has been
established for long time
\citep[e.g.,][]{2003ApJ...589L..21M,2004ApJ...604L..89H}. This implies
that the galaxy bulge and the black hole grow together and is one of the
key clues in the development of our understanding of the nature of 
galaxy formation. It is believed that cold gas is the major common 
source to fuel both star formation (SF) and central supermassive 
black hole growth. Strong active galactic nuclei (AGNs) are usually 
found in galaxies rapidly forming stars
\citep{2003MNRAS.346.1055K,2004ApJ...611L..85P,2004ApJ...613..109H,2005Natur.434..738A,2009MNRAS.397..135K,2009MNRAS.399.1907N,2009ApJ...696..891H,2010ApJ...712.1287L,2010A&A...518L..26S,2012A&A...540A.109S,2013ApJ...773....3C,2014ApJ...783...40G}.
The correlation between AGN activity and host SF is more tight in the
central regions of the galaxy than in the outskirts
\citep{2007ApJS..173..357K,2007MNRAS.381..543W}. In the local
Universe, around 0.1 percent of the total gas accreted onto the galaxy
finally falls into the central black hole \citep{2004ApJ...613..109H},
consistent with the observed $M_\mathrm{BH}/M_\mathrm{bulge}$ ratio.

There are several mechanisms, internal and external, to transport cold
gas from the outskirts of the galaxy, where the baryonic material is
not dense enough to collapse and to form stars, to the inner regions
of the galaxy. Galaxy interactions are known as one mechanism that can 
induce gas inflows, which can enhance the
SF in galaxies. There is observational evidence for this
\citep{2006AJ....132..197W,2007AJ....134..527W,2008AJ....135.1877E,2008MNRAS.385.1903L,2009ApJ...698.1437K},
as well as theoretical support from simulations
\citep{1983MNRAS.205.1009N,2000MNRAS.312..859S,2007A&A...468...61D}. This
naturally leads to a hypothesis that there is a tight connection
between AGN activity and galaxy-galaxy interactions
\citep{2007MNRAS.375.1017A,2011MNRAS.418.2043E,2012ApJ...745...94L,2013MNRAS.431.2661C}. However,
the observational evidence for this has been contradictory
\citep{2008MNRAS.385.1915L,2008AJ....135.1877E,2009MNRAS.397..623G,2010MNRAS.401.1552D,2011ApJ...743....2S,2012ApJ...744..148K,2013A&A...549A..46B}.

In this letter, we aim to test the AGN-interaction connection again
with a large, low-$z$ galaxy sample extracted from the Sloan Digital
Sky Survey \citep[\sdss,][]{2000AJ....120.1579Y}. In previous work,
also based on \sdss\ data, \citet{2008MNRAS.385.1915L} found that 
AGNs do not have more nearby companions compared with inactive
galaxies with the same level of SF in the host galaxy. However, these
authors used the narrow \oiii\ emission line as their AGN activity
indicator \citep[see e.g.][]{2004ApJ...613..109H}. The \oiii\ line is
emitted from the narrow line region, which is a few hundred parsecs
away from the central black hole. Although the \oiii\ line luminosity
is geometrically unobscured, it is an indirect measurement of the
black hole accretion rate. In this letter, we use the same neighbour
counting technique to quantify the environment of AGNs. Instead of the
\oiii\ line, we adopt another AGN activity indicator, the mid-infrared
(IR) emission from the torus, to study the AGN-interaction
connection. The nuclear mid-IR emission is from the very central part
of the galaxy with typical size of a few parsecs
\citep{2004Natur.429...47J,2007A&A...474..837T,2008A&A...486L..17B,2009ApJ...705L..53B,2009A&A...502...67T,2009A&A...493L..57K,2011A&A...536A..78K,2011A&A...527A.121K,2012ApJ...755..149H},
much smaller than the narrow line region. As was done in the first
paper of this series \citep[][hereafter Paper I]{2013MNRAS.436.3451S},
we use the data from the Wide-field Infrared Survey Explorer
\citep[\wise,][]{2010AJ....140.1868W,2011ApJ...731...53M} to estimate 
the strength of the torus emission.

In this letter we will discuss the relation between the environment of
local AGN host galaxies and their central black hole activity as
traced by their \oiii\ and IR luminosities. In section \ref{sec:data},
we will briefly describe the data we use in this letter. We will
present our results in section \ref{sec:results} and discuss them in
section \ref{sec:discuss}.

\section{Data}
\label{sec:data}

We start from a sample that includes all the galaxies from the MPA-JHU 
\sdss\ DR7 catalogue\footnote{http://www.mpa-garching.mpg.de/SDSS/} 
with $r$-band model magnitudes in the range $14.5<r<17.6$, stellar 
masses in the range $9.8<\log(M_\ast/\mathrm{M_\odot})<11.8$ and 
redshifts in the range $0.02<z<0.21$.
The \sdss\ catalogue provides reliable photometric and spectroscopic
data in this redshift and magnitude range for AGN host galaxies 
(see also Paper I), and is suitable for studying the environment of 
galaxies. The \sdss\ galaxies are matched to the \wise\ catalogue 
within a search radius of 3$^{\prime\prime}$ from the optical position. 
We use all the \wise\ detections with signal-to-noise ratio above 3.
96.8\% of the galaxies are detected in 4.6\mum\ band.
Optical classifications are based on the star-forming galaxy/AGN 
separation line suggested by \citet{2003MNRAS.346.1055K} on the BPT 
diagram \citep{1981PASP...93....5B}. Here we construct a sample of
18727 Seyfert galaxies (hereafter S1 sample). They are separated 
from LINERs (low ionization nuclear emission-line regions) according 
to the \sii/\ha\ ratio using Function 7 in the paper by
\citet{2006MNRAS.372..961K}. LINERs are excluded from the
analysis. Most of them have nuclei too faint to be detected in the
mid-IR, as discussed in Paper I. Some recent studies also suggest that
the \oiii\ emission in LINERs may have non-nuclear origin
\citep{2011MNRAS.413.1687C,2012ApJ...747...61Y}. In this letter, we
limit our sample to Seyfert galaxies to avoid these problems.

We use two different ways to quantify the AGN activity: optical
emission line and IR torus luminosities. The optical indicator is the
\oiii\ line luminosity \citep{2004ApJ...613..109H}, corrected for SF
contribution and dust extinction. The SF contribution fraction to 
the \oiii\ luminosity is estimated based on the position of
the galaxy on the BPT diagram, as suggested by
\citet{2009MNRAS.397..135K} (see their Figure 3). The extinction is
estimated from the Balmer decrement, assuming an intrinsic
\ha/\hb\ ratio of 2.87 for star-forming galaxies and 3.1 for AGN, and
the reddening curve from \citet{2007MNRAS.381..543W}.

The IR AGN activity indicator, nuclear IR luminosity, is
derived by subtracting the host contribution from the observed IR
luminosity (hereafter ``AGN/nuclear IR luminosity'' always refers to 
the host-subtracted IR luminosity). We match the AGN host galaxies 
to a control samples of non-AGNs with similar redshifts, stellar masses 
and star formation rates. We estimate the IR luminosities of the host 
component by averaging the IR luminosities of matched non-AGNs.  
This means that statistically, the resulting host-subtracted IR 
luminosities are dominated by nuclear emission. This subtraction 
technique is described in detail in Paper I.

We note that the PSF (point spread function) size 
of the \wise\ images is larger ($\sim6^{\prime\prime}$ at 4.6\mum) 
than the \sdss\ fiber aperture ($3^{\prime\prime}$). The AGN IR 
luminosities may thus be overestimated when there are close
companions unresolved by \wise. In this letter, we adopt the 
\wise\ profile-fit magnitude, which is the least sensitive to 
blending by nearby sources, in order to minimize this effect.
The profile-fitting routine uses both a passive and an active 
deblending technique to identify blended 
objects\footnote{http://wise2.ipac.caltech.edu/docs/release/allsky/expsup/sec4\_4c.html\#wpro}.
It is possible to deblend close neighbours with very small 
angular separations. The profile-fit magnitude is slightly different
from the elliptical aperture magnitude we use in Paper I,
but we find that for isolated galaxies this change has little effect 
on our AGN IR luminosity estimates.

In order to make sure that the unresolved galaxies do not affect our
results, we build a ``cleaner'' sample for comparison. For each galaxy
in the S1 sample which has close companions in \sdss\ with angular 
separation smaller than $6^{\prime\prime}$, we match the deblended 
\wise\ sources with the \sdss\ catalogue. If there is any close 
companion not successfully deblended, we discard the galaxy. In total 
we drop 551 galaxies from S1. This slightly trimmed sample contains 
18176 Seyfert galaxies (hereafter S2 sample). Their nuclear IR 
luminosities are more reliable, but the sample may be slightly less 
representative of the general population of Seyfert galaxies.

The black hole mass is estimated from the measured
stellar velocity dispersion of the galaxy by using the empirical
$M_\mathrm{BH}$--$\sigma$ relation provided by
\citet{2009ApJ...698..198G}. We use the 4000 \AA\ break \dfour, which
is relatively free from extinction effects, to estimate the age of the
central stellar population in AGN hosts \citep{2003MNRAS.341...33K}.
The reader is also referred to Paper I for more technical details 
about the data processing.

\section{Results}
\label{sec:results}

We calculate the average number of close neighbours in the vicinity 
of AGNs using the method described by \citet{2006MNRAS.373..457L}. 
We count the number of galaxies in the \sdss\ photometric sample 
brighter than a fixed $r$-band magnitude limit, within a given 
value of the projected radius $R_p$. We make a statistical correction 
for chance projections of foreground and background galaxies that 
lie along the line-of-sight, by calculating the counts around 
randomly placed points within the region of the sky covered by 
the \sdss\ DR7. The random positioning simulations and 
foreground/background subtraction are repeated for 50 times. 
From these realizations, we estimate the uncertainty in the 
calculated neighbour counts.

\begin{figure*}
  \centering \includegraphics[width=16cm]{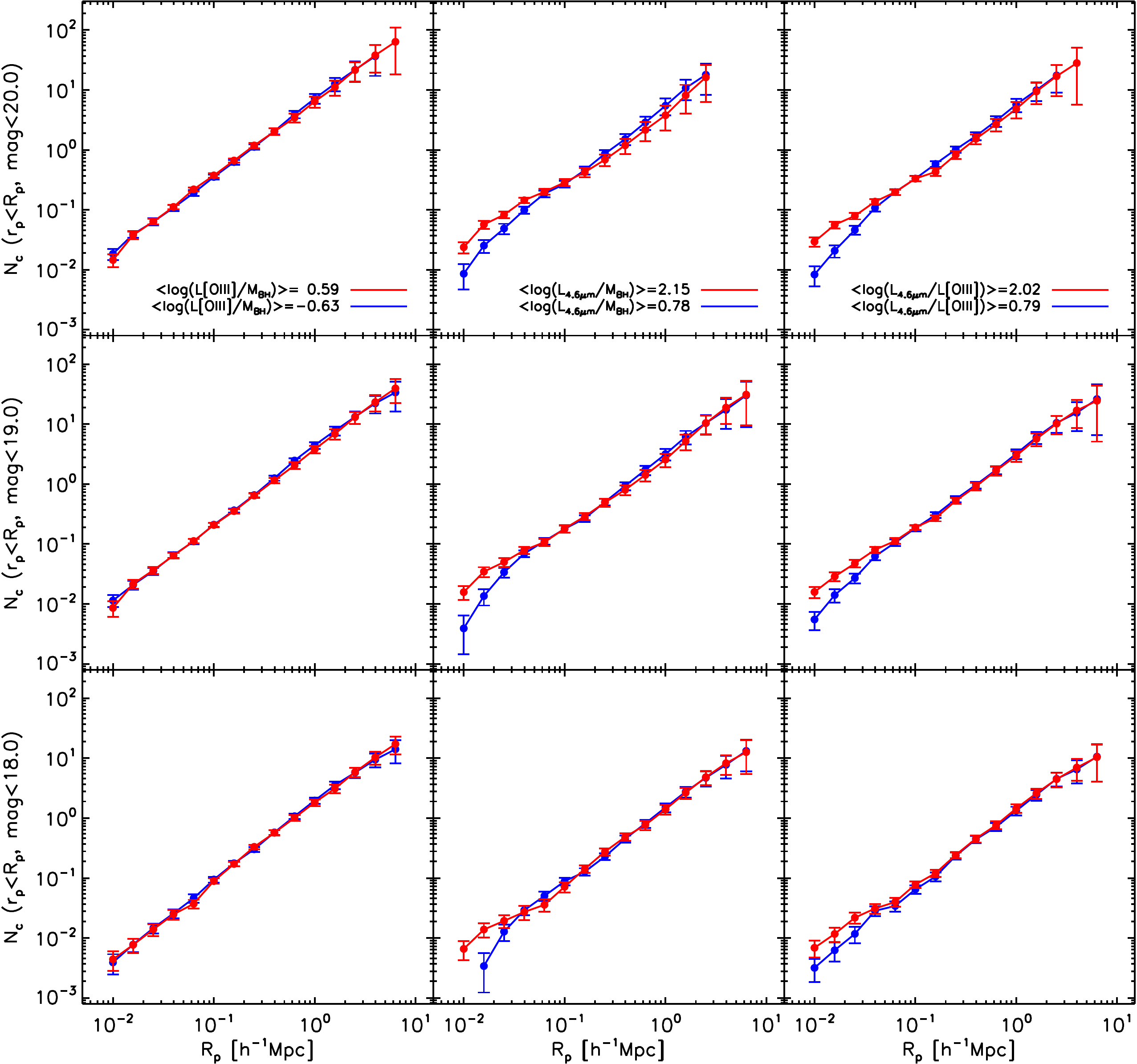}
  \caption{The average number of neighbours within a given 
    value of the projected radius for S1 galaxies. Top, middle and 
    bottom panels are for different magnitude limits of the \sdss\ 
    photometric sample used for the neighbour counting. Left, middle 
    and right panels are Seyferts binned by different \eddpar, 
    \wtwoeddpar\ and \wtworat, respectively. The results from 
    different bins are displayed in different colors, as shown in 
    the legend.}
  \label{fig:counts}
\end{figure*}
   
Figure \ref{fig:counts} shows the background-subtracted average number
of close neighbours within projected radius $R_p$. The results are for
different limiting magnitudes for the neighbour galaxies. In the
top, middle and bottom panels, the apparent magnitude limits for the
neighbours are 20.0, 19.0 and 18.0, respectively. From top to bottom,
brighter magnitude limits mean only the brighter (and statistically
more massive) neighbours are considered in the counting analysis.

In each panel, the Seyfert galaxies are split into two subsamples.
They are binned according to \eddpar\ in the left, \wtwoeddpar\ in 
the middle and \wtworat\ in the right panels, respectively. 
The galaxies are sorted according to the parameter in question, 
and the red and blue subsamples are extracted from the top 33\% 
and bottom 33\% of the objects, respectively. The two subsamples 
are then trimmed to make them closely matched in redshift, stellar 
mass, concentration and 4000 \AA\ break with the following 
tolerances: $\Delta z<0.01$; $\Delta(\log(M_*))<0.1$; 
$\Delta(R_{90}/R_{50})<0.2$ and $\Delta\dfour<0.05$. In this way, 
we make sure that the two subsamples of Seyfert galaxies with 
very different AGN activity properties are compared at the same 
redshift, with the same host stellar mass distributions, with
the same Hubble type distribution and with the same distribution 
of host SF activity. As mentioned in the introduction, it is known 
that the powerful optical/IR/X-ray AGNs are linked with SF, while 
the SF is also enhanced by galaxy interactions. It is important to 
compare different AGN subsamples with the same host properties, to 
make sure the results probe the effect of nuclear activity rather 
than SF in the host.

In the left panels, we can see that the number of neighbouring
galaxies has no dependence on AGN activity level measured by the
\oiii\ line emission, consistent with the results shown by
\citet{2008MNRAS.385.1915L}. At scales larger than 100 kpc, this is
still true if we compare AGNs with different \wtwoeddpar\ or
\wtworat. However, at scales smaller than 100 kpc, there is a
neighbour count excess for AGNs with higher \wtwoeddpar\ or
\wtworat. The excess signal is stronger when the projected
separation is smaller. For closer neighbours, the excess neighbour 
count is stronger. We do not see significant dependence of the 
neighbour counts on the limiting magnitude of the photometric sample. 
This implies that the excess signal is mainly contributed by
the brightest companions.

\begin{figure}
  \centering \includegraphics[width=8cm]{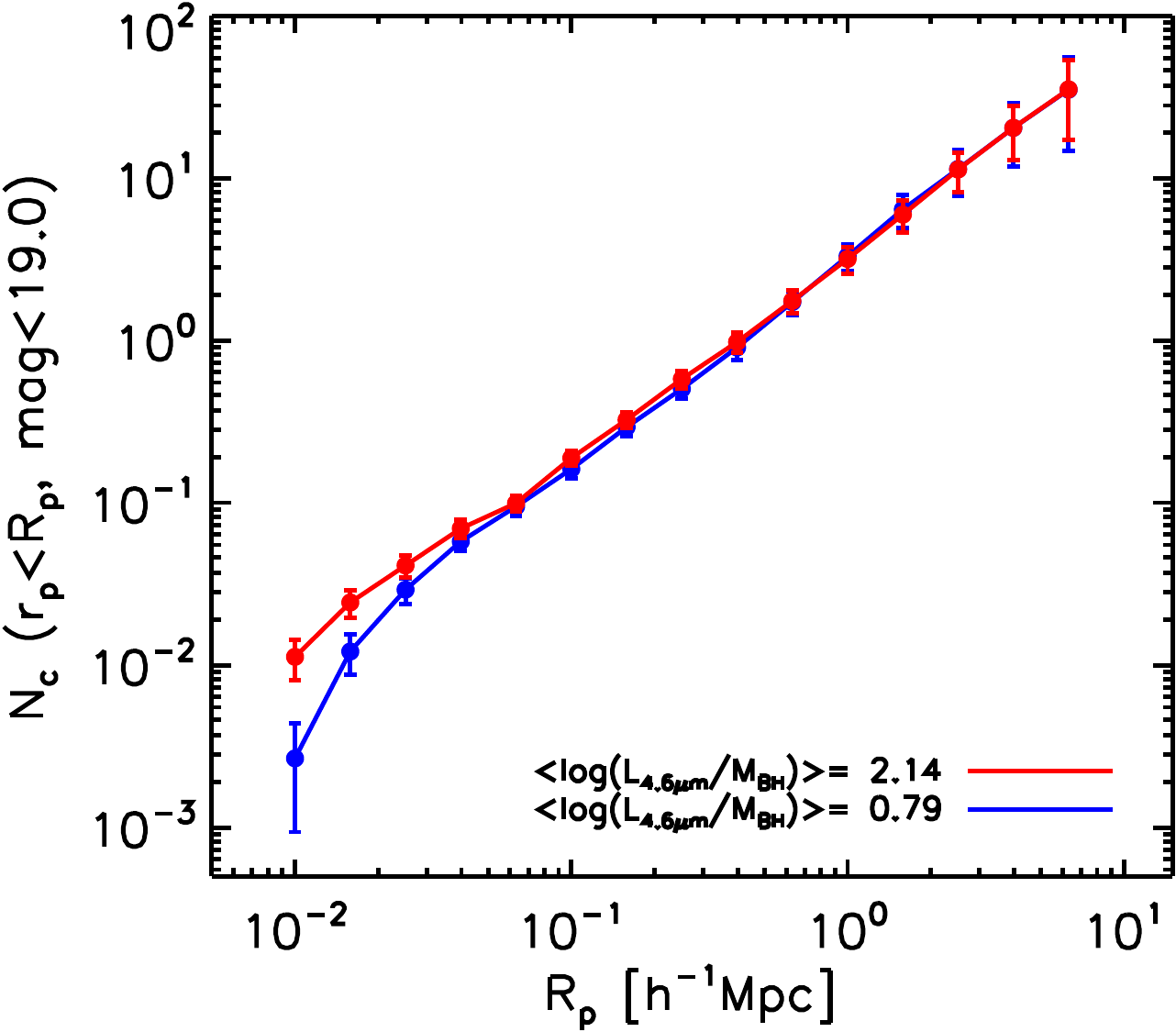}
  \caption{The same as the central middle panel of 
    Figure \ref{fig:counts} but using S2 sample.}
  \label{fig:more}
\end{figure}

We perform the same analysis on the S2 sample, and find the same
results as shown in Figure \ref{fig:more}. This suggests that
our results are not significantly affected by the neighbouring 
galaxies unresolved by \wise.

We also perform the same analysis on the S1 sample using the 22\mum\
band, another \wise\ band sensitive to AGN activity (see Paper I). 
Interestingly, we find the same trend as seen in 4.6\mum\ band, i.e. 
the AGNs with higher \wfoureddpar\ or \wfourrat\ have more close 
neighbours. This seems to further support the robustness of our result. 
However, ``cleaning'' unresolved neighbours in 22\mum\ band is 
much more difficult because the 22\mum\ PSF is twice larger than the 
4.6\mum\ one. If we perform the same ``cleaning'' procedure as 
we did to build S2 sample (of course we need to check the \wise\ 
detections in a radius of $12^{\prime\prime}$ around primary 
galaxy instead of $6^{\prime\prime}$), we find that too few galaxies 
with close neighbours are left. This prevents us from coming to a fully  
solid conclusion about the neighbour excess in the 22\mum\ band.

\section{Discussion}
\label{sec:discuss}

Theoretical simulations have suggested that merging is an efficient 
way to trigger nuclear activity in galaxies. It is believed that 
there is an early obscured stage tightly linked with the merging event
\citep{1991ApJ...370L..65B,2005Natur.433..604D,2006ApJS..163....1H}.
These AGNs are characterized by their prominent IR emission. One typical
well-known example is ultra-luminous infrared galaxies (ULIRGs) 
with hidden active nuclei \citep{1988ApJ...325...74S}. The 
merger-driven ULIRG-like objects are usually the most massive 
galaxies with extreme star formation rates. Compared with such objects,
our \sdss\ AGNs are less massive and usually have moderate star 
formation rates. It is thus very interesting to find an  
AGN-environment connection for these AGNs of moderate luminosity
and star formation rate. As shown in Figure \ref{fig:counts}, 
the link only exists when we use the host-subtracted IR 
luminosity as our AGN activity indicator.

In a recent paper, \citet{2014MNRAS.441.1297S} 
report a similar result. They start from a sample of \sdss\ mergers, 
showing that there are more \wise\ color-selected AGNs in close 
pairs and the AGN fraction is higher at smaller projected separation. 
This is consistent with our results.

Our results mean that it is necessary to understand the 
difference between \oiii-bright and IR-bright AGNs. Both our 
\oiii\ and IR luminosities are corrected for the contribution from 
the host galaxy. The \oiii\ luminosity is also extinction corrected. 
In practice, the IR-to-\oiii\ luminosity ratio probes the relative 
contribution of AGN narrow line region emission and AGN IR torus
emission, free from host contamination. Physically, the IR
luminosity from the torus reflects the obscured part of the total 
AGN radiation power, while the \oiii\ luminosity is an indicator 
of the unobscured part. The IR-to-\oiii\ ratio is hence a direct 
probe of the torus covering factor, and is directly related to the 
thickness of the torus. It implies that in our sample the difference 
between optical and IR AGN indicators is likely caused by different 
torus structures in different galaxies.

As mentioned in the introduction, it is commonly accepted that 
galaxy interactions can boost the inflow of cold gas. In a recent 
theoretical paper, \citet{2012MNRAS.420..320H} suggest that the 
torus can form as a result of cold gas inflows. The torus grows 
thicker, forming warp-like structures, due to inflow-induced 
instabilities. Based on this, it is not surprising to find a  
torus-interaction link. Further investigations with better 
observational data and better simulations are necessary to make 
further progress in understanding these issues.

The merger-driven evolutionary scenario suggests a time
lag between AGN IR and optical luminosity peaks
\citep[e.g.][]{2008ApJS..175..356H}. The difference between 
\oiii-bright and IR-bright AGNs shown in Figure \ref{fig:counts}
also implies that IR-bright AGNs may be seen before the merging 
event and \oiii-bright AGN afterwards. In Paper I, we stated that           
the host galaxy properties of IR and optical AGNs do not show 
significant differences. For our \sdss\ Seyfert galaxies, it seems 
that the total stellar mass and black hole mass accumulated during 
the obscured stage are negligible. We conclude that either only a 
small fraction of the local Seyfert galaxies are triggered by mergers,
or the obscured phase is very short compared with the whole AGN lifetime.

In summary, in this letter, we have used a neighbour 
count technique to study a large sample of local Seyfert 2 
galaxies. We have found clear evidence that the IR-strong 
AGNs are tightly connected with galaxy interactions. 
AGNs with higher nuclear IR luminosities are more likely to be 
surrounded by companion galaxies within a projected distance 
of less than 100 kpc. The strength of the neighbour count excess increases 
with decreasing projected distance.

\section*{Acknowledgments}

The work is sponsored by the exchange program between Max Planck
Society and Chinese Academy of Sciences (CAS).
CL acknowledge the support of National Key Basic Research Program 
of China (No. 2015CB857004), NSFC (Grant No. 11173045, 11233005, 
11325314, 11320101002) and the Strategic Priority Research 
Program ``The Emergence of Cosmological Structures'' of CAS 
(Grant No. XDB09000000). This work has made use of data from 
the \sdss, \sdss-II, \wise\ and \textit{NEOWISE}.

\bibliographystyle{mn2e}
\bibliography{agn}

\label{lastpage}

\end{document}